# Plastic forming of metals at the nanoscale: interdiffusion-induced bending of bimetallic nanowhiskers

Yuanshen Qi[1*], Gunther Richter[2], Eylül Suadiye[2], Michael Kalina[1], Eugen Rabkin[1*]

[1]Department of Materials Science and Engineering, Technion – Israel Institute of Technology, 3200003 Haifa, Israel

[2]Max Planck Institute for Intelligent Systems, Heisenbergstrasse 3, 70569 Stuttgart, Germany

*Correspondence to: yuanshen.qi@campus.technion.ac.il; erabkin@technion.ac.il

Address: Department of Materials Science and Engineering, Technion – Israel Institute of Technology, 3200003 Haifa, Israel

**Abstract:**

Controlled plastic forming of nanoscale metallic objects by applying mechanical load is a challenge, since defect-free nanocrystals usually yield at near theoretical shear strength, followed by an uncontrolled catastrophic failure. Herein, instead of mechanical load, we utilize chemical stress from imbalanced interdiffusion to manipulate the shape of nanowhiskers. Bimetallic Au-Fe nanowhiskers with an ultra-high bending strength were synthesized employing the molecular beam epitaxy technique. The one-sided Fe coating on the defect-free, single-crystalline Au nanowhisker exhibited both single- and poly-crystalline regions. Annealing the bimetallic nanowhiskers at elevated temperatures led to gradual change of curvature and irreversible bending, which is attributed to the grain boundary Kirkendall effect during the diffusion of Au along the grain boundaries in the Fe layer. The results of this study demonstrate a high potential of chemical interdiffusion in the controlled plastic forming of ultra-strong metal nanostructures.

**Main:**

Plastic deformation of polycrystalline bulk metallic materials at ambient temperature is mediated by the glide of pre-existing lattice dislocations and generation of new dislocations by the pre-existing and newly formed dislocation sources [1], formation of stacking faults and deformation twinning [2], grain boundary sliding [3] or phase transformations [4], depending on the grain size, stacking fault energy, strain rate, composition, etc. The typical elastic deformation of bulk metallic materials at the onset of plastic yield is about 0.1%. In contrast, plastic deformation of single-crystalline nearly defect-free metallic nanocrystals such as nanowhiskers/nanowires (NWs) [5] and nanoparticles [6] is controlled by the nucleation of new dislocations. In the uniaxial deformation regime, the nanocrystal deforms elastically up to very high strains of several percent and stresses of several Giga Pascals (GPa), followed by a catastrophic plastic collapse [6]. This deformation mode is referred to as nucleation-controlled plasticity [7-9].

The nucleation-controlled plasticity of nanocrystals, on one hand, leads to high mechanical strength approaching its upper theoretical limit [6-11]; on the other hand, it makes the manipulation of the crystal morphologies by plastic-forming process very difficult [12-14] (see Supplementary Movie 1). Crikor et al. demonstrated that the intermittent dislocation avalanches in microcrystals under loading result in stochastic distribution of plastic strain [12]. For this reason, the mechanical loading of the defect-free metal nanocrystals cannot be employed for their controlled plastic forming into a desired shape. Such forming may be necessary because the variety of shapes of the as-synthesized defect-free nanocrystals is severely limited by the relative specific surface energies of the crystal facets [15, 16].

Controlling of the morphology of NWs is critical for their potential applications such as plasmonic waveguide [17-19]. Studies of mechanical behavior of NWs employing three-point bending and cantilever beam bending reveal high elastic strains at the onsets of plastic yielding, often followed by a strain burst and abrupt fracture [20-23]. Currently, the curvature or morphology of the metal NWs cannot be precisely manipulated by plastic deformation via mechanical load [24].

Apart from the stress generated by mechanical load, stresses induced by chemical interdiffusion represent an alternative route of plastic deformation [25]. For example, the imbalance of atomic diffusion fluxes during the Kirkendall effect results in net vacancy flux and concomitant lattice drift and shape changes due to the climb of edge dislocations [25-28]. The scarcity of internal vacancy sinks may result in the generation of high internal elastic stresses and viscous material flow [25, 29]. In fact, in the past decade Kirkendall effect has been widely utilized in the synthesis of hollow metallic nanostructures [30-32]. However, the possibility of using the interdiffusion-generated internal stress for controlled plastic forming at the nanoscale has not yet been explored. Herein, we performed annealings of Au/Fe bimetallic NWs to demonstrate the feasibility of controlled plastic forming of NWs by interdiffusion-generated stresses.

The Au/Fe bimetallic NWs were prepared by physical vapor deposition onto W substrates using molecular beam epitaxy (MBE), under a near-equilibrium growth condition. The as-grown faceted single-crystalline <011>-oriented Au NWs (of ~360 nm in cross-sectional width and of several micrometers in length) were coated on one side with Fe layers of ~200 or ~50 nm in thickness in the same MBE chamber, without breaking the ultra-high vacuum.

One typical as-synthesized bimetallic NW is presented in Fig. 1(a). The closest-packed {111} and second closest-packed {100} planes of the face-centered cubic lattice form the side facets of Au NWs. The Fe layers deposited on the {100} and {111} facets were found to be single-crystalline (SX) and poly-crystalline (PX), respectively, and to form the coherent SX Fe – Au and incoherent PX Fe - Au interfaces (see Figure S1 in Supplementary material). The coherent interface exhibits a clear Bain orientation relationship between the SX Fe layer and the Au NW, e.g. $[0\bar{1}1]_{Au}//[001]_{Fe}$, $(011)_{Au}//(010)_{Fe}$ and $(100)_{Au}//(100)_{Fe}$. It is worth noting the presence of grain boundaries (GBs) and nano-cavities in the PX Fe layer, and at the interface between PX and SX Fe layers (Fig. 1b and Figs. S2-S5). We used the easy-lift technique in a focused ion beam – scanning electron microscope (FIB-SEM) dual beam instrument to harvest and transfer individual NWs from the W substrate to Mo foil substrates, see Fig. 1(c, d).

The as-synthesized bimetallic NW was slightly bent in the direction towards Fe coating (Fig. 1d), which contrasts with the straight morphology of single-phase Au as well as Cu and Pd NWs prepared using same MBE procedures [5]. This bending can be explained by a combined effect of lattice mismatch strain and islands-coalescence stress in the SX and PX Fe layers, respectively (for the estimate of their contributions see Supplementary material and Figure S2). The lattice mismatch strain at the SX Fe – Au interface, between $(011)_{Au}//(010)_{Fe}$ can be estimated as $\varepsilon_m = (d_{Fe(010)} - d_{Au(011)})/d_{Au(011)} = -0.0059$, where $d_{Fe(010)} = 2.867$ Å and $d_{Au(011)} = 2.884$ Å are the interplanar spacings in Fe and Au, respectively [33]. The radius of curvature associated with this mismatch strain was estimated to be 40 µm employing the Timoshenko formula [34]. The maximum tensile stress generated by the islands zipping during the Volmer-Weber growth of PX Fe layers was estimated to be 19.1 GPa [35]; this tensile stress also contributes to the bending of NW towards the Fe layer. However, it is worth noting that with increasing film thickness, the islands-coalescence tensile stress is gradually relaxing [36], also by plastic deformation of the Au NW resulting in the formation of ledges at the PX Fe-Au interface, see Figure S3.

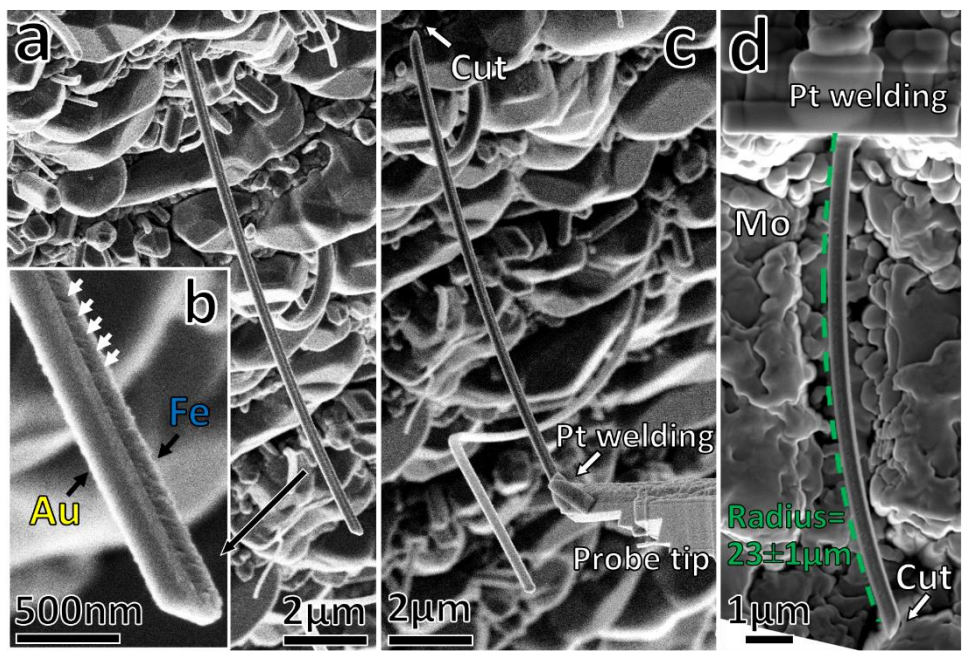

**Figure 1. Au/Fe bimetallic NW characterization, harvesting and mounting**. Secondary electron (SE)-SEM micrographs of (a) a NW grown on a W substrate; insert (b) shows an enlarged view to distinguish the Fe layer on the Au NW; white arrows mark some of the morphological features on the polycrystalline Fe layer associated with the GBs or nano-cavities between the Fe nanocrystals formed by the Volmer-Weber growth mechanism; (c) welding of the probe tip and the NW employing Pt deposition followed by cutting of the NW free from the W substrate; (d) mounting of the NW on a Mo foil substrate by Pt deposition and cutting it free from the probe tip; the green dashed line highlights the radius of curvature of the NW. The error in radius of curvature is related to the measurement uncertainties.

Three NWs were lifted-out, heat treated and characterized in this study. NW#1 and NW#2 with a ~200 nm thick Fe layer were employed in *in-situ* annealing experiments which were conducted in a Zeiss Ultra-Plus SEM. These two NWs were mounted on a Mo foil which was fitted into the heating stage, and the annealing was performed under the vacuum of $5\cdot 10^{-5}$ mbar in the SEM chamber. Before each SEM image acquisition, the temperature was kept constant for 20 min for thermal equilibration and stage stabilization. The NW#3 with a ~50 nm thick Fe layer was annealed *ex-situ*.

During the *in-situ* heating of NW#1, we firstly observed a reversible bending behavior at relatively low temperatures, see Fig. 2(a-c). The radius of curvature, *R*, changed from 23±1 µm at room temperature to 18±1 µm at 300 °C, increasing back to 21±1 µm upon cooling to 25 °C after the heating system was switched off. This reversible bending behavior was attributed to the elastic deformation due to the mismatch of the thermal expansion coefficients of the Au NW and the Fe coating ($14.2\cdot 10^{-6}$ °C$^{-1}$ and $11.4\cdot 10^{-6}$ °C$^{-1}$, respectively). Higher thermal expansion of Au compared to that of Fe upon heating from room temperature to 300 °C resulted in the development of compressive stress in the Au NW, as the mismatch strain rose from -0.0059 to -0.0067. This increase in the mismatch strain changes the estimated value of *R* from 40 to 36 µm, in qualitative agreement with the experimental results. It was also noticed in the experiment that the radius of curvature of the NW#1 at 300 °C has not fully recovered from 18 back to 23 µm upon cooling, since the measured value of *R* was 21 µm after the full heating/cooling cycle. This was due to the healing of some nano-cavities in the Fe layer at 300 °C, which resulted in volume shrinkage in the Fe coating (see Figures S4-S6 and estimation of nano-cavity healing based on Fe GB diffusion in Supplementary material).

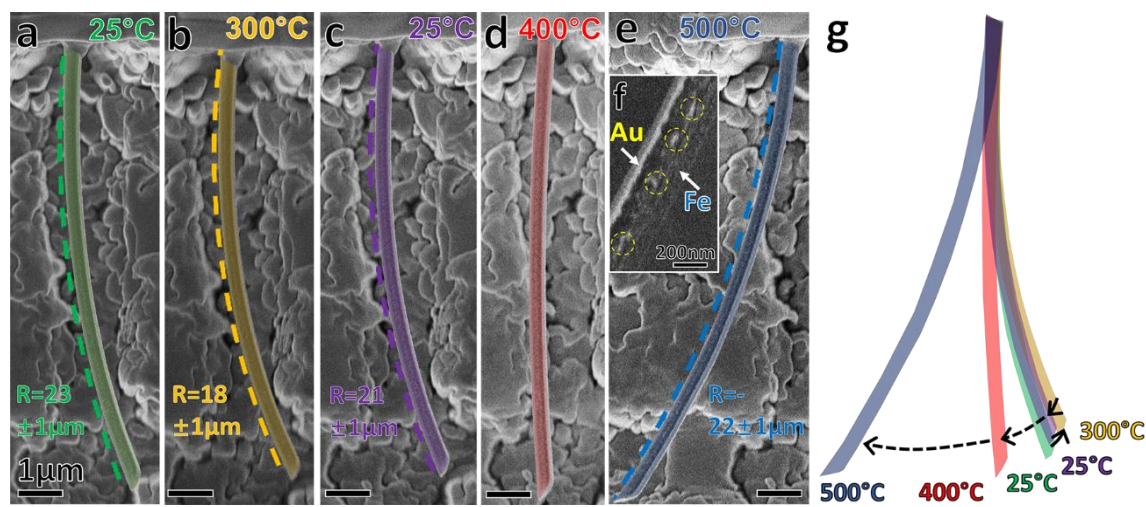

**Figure 2. Reversible and irreversible bending of a Au/Fe bimetallic NW at elevated temperatures**. SE-SEM micrographs of (a-c) reversible bending of NW#1 at 300 °C in elastic deformation regime; (c-e) irreversible bending of NW#1 at 400 °C and 500 °C. The insert (f) shows an enlarged view of the NW acquired using backscattered electrons (BSE), visualizing the Au clusters on the Fe layer by their Z-contrast. (g) illustrates schematically the bending process. The radii of curvature, *R*, of the NW at each temperature are presented in (a-e). The error in determining of the radius of curvature was estimated by performing 5-6 independent curvature measurements.

Irreversible bending in the opposite direction occurred when the temperature was increased above 300 °C. NW#1 was nearly straight at 400 °C; and kept bending towards Au NW reaching a curvature of 22±1 µm with an opposite sign at 500 °C, see Fig. 2(e). At this stage, the Fe coating and Au NW were under compression and tension, respectively, resulting in a gradual lattice rotation of Au NW about [011] zone axis, which is normal to the Au NW growth direction [0$\bar{1}$1] (see transmission Kikuchi diffraction (TKD) characterization in Figure S7 in Supplementary material). Finally, from the BSE-SEM micrograph shown in Fig. 2(f), some isolated Au clusters were observed on the outer surface of Fe layer, which indicated significant Au diffusion at elevated temperatures. Indeed, the diffusional penetration of Au along the GBs and nano-cavities in the Fe coating can explain the volume expansion and concomitant development of compressive stresses in the latter.

The Au flux penetrated the Fe coating via the nano-cavities and GBs, as demonstrated in Fig. 3. This penetration resulted in the formation of Au-rich clusters on the outer surface of Fe layer shown in Fig. 2(f), and highlighted in both normal (Fig. 3a) and longitudinal (Fig. 3d) cross-section views of high-angle annular dark-field scanning electron microscopy (HAADF-STEM) images of NW#1. The GB diffusion process led to the formation of characteristic GB grooves at the Au-Fe interface, which are marked in Fig. 3 (d, e). Also, the nano-cavities in Fe got partially filled with the Au-rich alloy. The diffusion of Au along, and segregation at the GBs in Fe nanostructures attached to the substrate has also been observed earlier by Amram et al [37]. These phenomena have been utilized in bulk Fe-Au alloys for filling and healing of the creep-induced micro-cavities at elevated temperatures, thus enhancing the alloy component lifetime [38].

As will be detailed below, Au diffusion along, and accretion at the GBs in the Fe layer caused its volume expansion, and concomitant bending of the NW in the direction of Au. At the same time, Fe also diffused into Au via lattice diffusion. As seen in Fig. 3(b), Au-rich phase grew at the expense of Fe, and the Au-Fe interface migrated in the Fe direction. This movement reflects high solubility of Fe in Au and negligible solubility of Au in Fe [39]. This volume interdiffusion and interface migration caused volume shrinkage of the Au constituent, since the lattice parameter of the Au-Fe alloys decreases with increasing Fe content. Finally, a twin boundary (TB) consisting of coherent and incoherent sections can be seen in Fig. 3(c). It nucleated on the free surface of Au NW, migrated towards the Au-Fe interface and stopped in the NW interior. This TB can be categorized as a deformation twin nucleated to relieve the diffusion-generated elastic stresses in the Au NW. The twin has nucleated on the free surface of Au NW rather than at the Au-Fe interface (the source of diffusion-generated stress) because the nucleation of the Shockley partial dislocations (also serving as twinning dislocations) is easier on the free surface than at the interface exhibiting lower diffusion mobility of Au atoms [40].

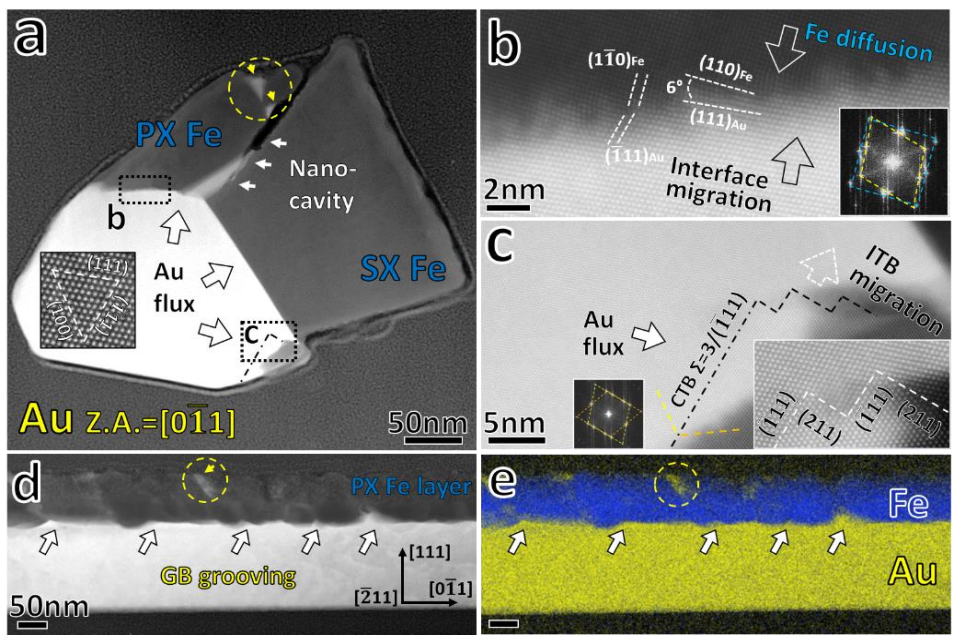

**Fig. 3. Plastic bending of the Au/Fe metallic NW induced by interdiffusion.** STEM-EDX characterization of bent NW#1 after *in-situ* heating at 500 °C. (a-c) HAADF-STEM images of the normal cross-section of NW#1, imaged along the [0$\bar{1}$1] zone axis (Z.A.) of Au. The diffusion flux of Au atoms into a Fe nano-cavity at the intersect of Au (111) and (100) facets is schematically shown by an arrow. (b) nano-roughness developed at the incoherent Au-Fe interface facilitating interface migration and Fe diffusion into Au; insert fast Fourier transform (FFT) image and lattice analysis indicate the incoherency of the interface; (c) a coherent twin boundary (CTB) and incoherent twin boundary (ITB) developed in Au to relax the diffusion-induced stress; (d, e) HAADF-STEM micrograph and corresponding EDX mapping of the longitudinal cross-section of NW#1, imaged along the [$\bar{2}$11] Z.A. of Au.

The interdiffusion induced plastic bending was reconfirmed in the bent NW#2, which went through the same *in-situ* heating cycle up to 500 °C as NW#1, see Fig. 4 (a, b). Au diffusion via Fe GBs and nano-cavities, and Fe lattice diffusion into Au and accompanying interface migration are shown in Fig. 4 (c, h, i). These micrographs support the hypothesis that the plastic bending has resulted from the interdiffusion between the Au NW and Fe coating.

Furthermore, a <011> asymmetric tilt GB characterized as a special $\Sigma=43/(455)_1/(\bar{5}33)_2/99.37°$ GB was found at the "mouth" of Au layer protruding into the gap between SX and PX Fe layers, Figure 4 (c, d). The formation of the GB could be understood in terms of the reduction of the energy of all GBs and interfaces in the system, see Figure S8 and the estimate in Supplementary material. The penetration of Au into the gap between two Fe layers in homoepitaxial orientation relationship with the Au NW would result in the (277) plane of Au abutting the SX Fe layer. Such Au-SX Fe interface exhibits high energy. Instead, Au penetrated the gap between the two Fe layers in a different orientation (rotated about $[0\bar{1}1]$ axis by 99° with respect to original Au NW), resulting in low-index (100) plane of Au contacting the SX Fe layer at the interface, and reducing the energy of the Au-SX Fe interface [41]. The energy "penalty" of this process is the energy of the Σ43 GB formed at the "mouth", yet because its total length is significantly smaller than the length of the Au-SX Fe interface the process is energetically favorable. Finally, three deformation TBs were observed, extending from the free surface to the Au-Fe interface across the width of the Au NW, see Figure 4 (c, f) and Figure S9. It is worth noting that unlike the TBs formed in Au NWs during tensile testing experiments, which are inclined with respect to the NW growth axis [42], the TBs observed here are parallel to the growth axis and extend over the whole length of the NW. This is due to difference in the stress states and Schmid factors for the {111}[011] slip system during uniaxial loading [42] and diffusion-induced bending uncovered in the present work.

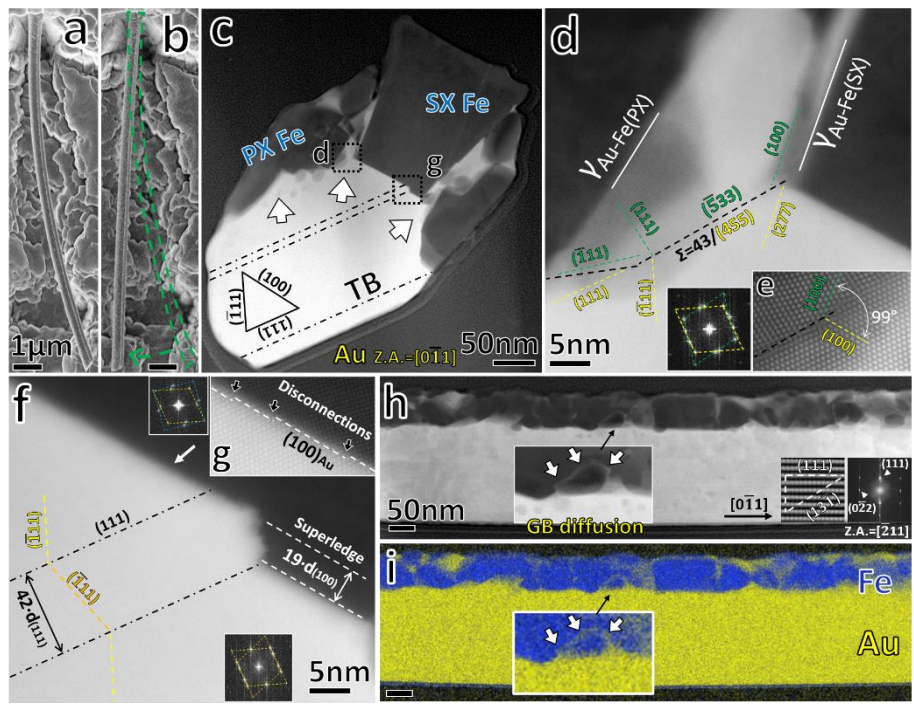

**Figure 4. Plastic bending of the Au/Fe metallic NW mediated by tilt GBs.** SEM and STEM-EDX characterization of bent NW#2 after *in-situ* heating at 500 ºC. (a, b) SE-SEM images showing the bending of NW#2 at 500 ºC, (c-g) HAADF-STEM images of the normal cross-section of NW#2, acquired along the $[0\bar{1}1]$ Z.A. of Au, (d, e) a tilt GB formed at the mouth of Au layer penetrating the gap between the PX and SX Fe layers, (f) deformation twins formed to relax the interdiffusion-induced stresses and to mediate plasticity, (g) disconnections and a superledge formed at the coherent Au-Fe interface indicating the interface migration towards Fe, (h, i) HAADF-STEM micrograph and corresponding EDX mapping of the longitudinal cross-section of NW#2, imaged along the $[\bar{2}11]$ Z.A. of Au; a Fe grain fully covered with Au GB segregation layer is highlighted.

The irreversible bending of NW#1 and NW#2 caused by *in-situ* heating, together with the STEM-EDX and TKD characterization of the bent NWs indicate that the diffusion of Au into Fe nano-cavities and GBs caused the lateral (i.e. parallel to the NW axis) expansion of the Fe layer. Let us first assume that the partial diffusion coefficients of Au and Fe along the Fe GBs are equal. In this case, all Au atoms penetrating along the Fe GBs will replace the Fe atoms there, and, with the average grain size in the PX Fe of 66 nm, one monolayer of Au at the GB will cause a lateral strain of $6\times10^{-4}$, one order of magnitude lower than the initial mismatch strain. However, in the case Au diffuses along the GBs much faster than Fe (GB Kirkendall effect [43]), the accretion of excess Au at the GB will not be accompanied by any outdiffusion of Fe. In this case, accretion of one monolayer of Au at the GBs will cause a Kirkendall strain of 0.004, comparable with the lattice mismatch strain (see Figure S10 and the estimation in Supplementary material). Formation of the GB diffusion wedge of 2-3 Au monolayers in thickness can fully compensate the initial mismatch strain, and cause the NW bending of similar magnitude in the opposite direction. Therefore, the underlying mechanism of the plastic bending is the imbalanced GB interdiffusion in the PX Fe layer and the GB Kirkendall effect [43]. Moreover, Fe lattice diffusion into Au contributed to the lateral shrinkage of the Au NW, because the lattice parameter of the Au(Fe) alloys decreases with increasing Fe content [39], and partial lattice diffusivities of Au and Fe in dilute Au(Fe) solid solution are nearly equal [44]. This shrinkage contributes to the reversal of the NW curvature at high temperatures. However, we would like to emphasize that at relatively low temperatures and short annealing times employed in the *in-situ* heating of NW#1 and NW#2, lattice diffusion of Fe into Au is very limited (as evidenced by the limited interface migration distance and formation of disconnections at the interface), therefore, the GB Kirkendall effect is the dominant factor in the plastic bending of the NWs when annealing was conducted below 500 °C.

Furthermore, our TKD measurements of the longitudinal cross-sections of the Au NWs did not reveal any low-angle GBs. According to the classical polygonization mechanism, it was expected that geometrically necessary dislocations (GNDs) would accommodate the plastic bending and self-organize at elevated temperatures in the form of low-angle GBs [45]. In contrast, we observed a gradual change of lattice orientation and formation of TBs in the bent Au NWs (See Figures S7, S9). This is consistent with the defect-free nature of as-grown Au NWs (no pre-existing dislocations or dislocation sources), and indicates that the energy barrier associated with nucleation of twinning dislocations (Shockley partials) is lower than that of GNDs.

We also tested the feasibility of using a thin layer of Fe to bend the Au NW and then dissolve the Fe. We employed the NW#3 with a ~50 nm thick Fe layer for this experiment. The NW was annealed in a rapid thermal annealing furnace under the reducing gas flow (Ar-10% $H_2$, 6N purity) at the temperature of 600 °C for 30 min. As can be seen in Fig. 5 (a-d), the radius of curvature changed from 43 to -56 µm after annealing, with most of the Fe layer being dissolved in the Au NW.

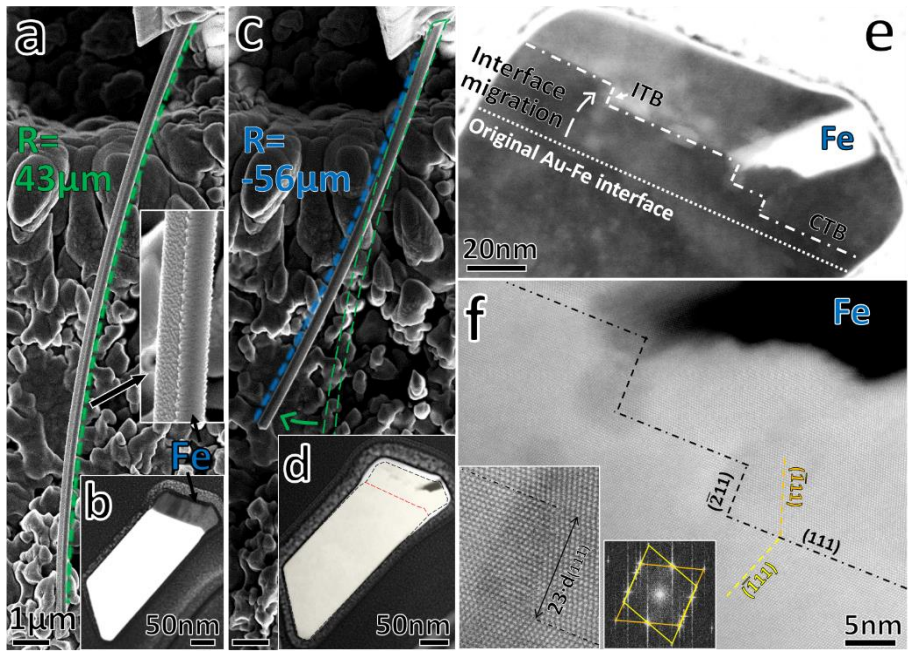

**Fig. 5. Bending by dissolving Fe layer in the Au NW,** revealed by the characterization of NW#3 after annealing at 600 ºC for 30 min. (a-d) the morphology and cross-section of the as-synthesized and annealed NW#3, (e) a dark-field (DF)-STEM image showing the TBs formed in the interdiffusion zone during the bending and interface migration, (f) a HAADF-STEM image characterizing the TB.

We estimated the volume fractions of the Fe layer and the Au NW in the pristine bimetallic NW to be 16.5% and 83.5%, respectively (Figure S11). This is equivalent to the overall composition of Au-13.5 at% Fe, so that full homogenization of the NW would result in a decrease of lattice parameter from 4.078 Å (in pure Au) to 4.041 Å [39, 46]. Taking into account that the lattice parameter of BCC Fe is 2.867 Å, full homogenization of the NW would result in a volume swelling of 6.5%. This estimate is very close to the measured increase of the cross-sectional area of 5.5% determined by comparing Figs 4b and d. This indicates that even nearly full dissolution of the Fe layer does not cancel internal stresses in the NW and its plastic bending. This is because the Au NW contracts upon dissolving Fe (due to the decrease of the lattice parameter of the Au(Fe) solid solution), while the former Fe coating expands due to its transformation into the Au(Fe) solid solution (with the atomic volumes of 11.78 Å$^3$ and 16.50 Å$^3$ in BCC Fe and in the Au(Fe) solid solution, respectively). A closer observation showed a TB formed in the interdiffusion zone and exhibiting both CTB and ITB sections. It formed during the migration of the Au-Fe interface towards the Fe layer and may be responsible for partial relaxation of the bending stresses (Figure S12).

In conclusion, we have successfully demonstrated the feasibility of controlled plastic deformation of the ultra-strong metallic nanostructures through the chemical interdiffusion route – employing either the GB Kirkendall effect or full diffusion intermixing. By performing *in-situ* heating on Au-Fe bimetallic NWs, we systematically studied the evolution of NW curvature associated with lattice mismatch strain, island-coalescence stress, thermal expansion mismatch, imbalanced GB interdiffusion, and full diffusion intermixing. We uncovered that when the GBs represent the main diffusion paths the irreversible plastic bending is dominated by the Au accretion at the GBs in polycrystalline Fe (GB Kirkendall effect). At higher temperatures or longer annealing times, bulk interdiffusion controls the plastic bending through the dependence of the lattice parameter on the composition, and the volume effect associated with migrating interphase boundary. Therefore, the design of the thickness and the microstructure of the diffusion layer, as well as the annealing parameters enables the bending curvature of a bimetal nanowhisker to be fine-tuned. Conversely, the bimetallic nanowhiskers with a known microstructure can be utilized as local temperature nano-sensors, converting the temperature and thermal history of the sample into geometrical curvature.

**Methods:**

Au/Fe bimetallic nanowhiskers preparation

Au/Fe bimetallic nanowhiskers (NWs) were prepared by physical vapor deposition onto W substrates using molecular beam epitaxy (MBE), under a near-equilibrium growth condition. The Au NWs (with a cross-section width of ~360 nm and length of several micrometers) were grown along <011> direction at 800°C; followed by the deposition of Fe layers (with a thickness of ~200 nm and ~50 nm) at room temperature. Both depositions were carried out in the same chamber without breaking the ultra-high vacuum of $5 \cdot 10^{-10}$ mbar. The estimation of the NW diameter and Fe layer thickness was based on the reading of a quartz balance measuring the deposition rate. Rates were 0.05 nm/s and 0.01 nm/s for Au and Fe, respectively. The deposition angle of the Fe is 30° relative to the substrate normal. The topological features of the Fe layer varied depending on the inclination angle of the Au NWs. Moreover, due to the shadowing effect, nano-cavities in the Fe layers have formed, particularly at the edges between the side {111} and {100} facets of Au NW.

Harvesting nanowhiskers:

Au/Fe bimetallic NWs were harvested using an easy-lift system in a focused ion beam (FIB) - scanning electron microscope (SEM) dual beam instrument (FEI Helios Nanolab Dualbeam $G^3$). Then they were mounted on Mo foil substrates using Pt deposition for further heating experiments. Mo foils were used as substrates for transferring individual NWs for further heat treatments due to their high oxidation resistance, thermal conductivity, and strength [47]. The NWs can be cut into several segments for the comparison of the cross-section microstructures before and after the annealing treatments. To avoid Ga ion beam damage of the NWs, very low ion beam currents, short dwell time (50-100 nano seconds), and small pixel number ($738 \times 512$ or $1536 \times 1024$) were used for imaging (30 keV, 1.1 pA) and cutting (30 keV; 7 pA), respectively.

Annealing treatments:
The *in-situ* heating experiments were conducted in a Zeiss Ultra Plus high-resolution scanning electron microscope (HR-SEM) equipped with a heating stage (Kammrath Weiss heating module 1050 °C). A heating rate of 5 °C s$^{-1}$ was used for the annealing. The *ex-situ* heating experiments were conducted in a rapid thermal annealing furnace (RTA; ULVAC-RIKO MILA 5000 P-N) under the reducing gas flow (Ar-10% $H_2$, 6N purity). A heating rate of 40 °C s$^{-1}$ was used and fast cooling was performed by simply switching off the heating. To avoid contaminations, the Mo foil with mounted NW(s) was placed on a sapphire plate, which in turn was placed on the quartz holder.

Electron microscopy characterization:

The atomic resolution microstructure features and the composition profiles of the NWs were examined in a scanning transmission electron microscope (STEM). The NW cross-section TEM lamellas were prepared by FIB using standard procedures. STEM was performed with a double Cs-corrected FEI 80-300 Themis G$^2$ operated at 300 kV. A 21 mrad beam convergence semi-angle was used and the resolution better than 0.9 Å was achieved. In high-angle annular dark-field (HAADF-) STEM imaging mode, the camera length was set to 94 mm, giving an inner collection semi-angle of 119 mrad and an outer collection semi-angle of 200 mrad. The bright-field (BF) STEM and low-angle annular dark-field (LAADF) STEM images were also collected in the semi-angle ranges of 0~11 mrad and 19~33 mrad. Energy-dispersive X-ray spectroscopy (EDX) mapping was carried in STEM using a Dual-X detector (Bruker). TEM selected area diffraction patterns (SADPs) were collected on a FEI Technai T20 operating at 200 keV. Finally, on-axis transmission Kikuchi diffraction (TKD) measurements were performed in a Zeiss Ultra Plus SEM equipped with a Bruker TKD system.

As-synthesized Au/Fe bimetallic NWs

The closest-packed {111} and second closest-packed {100} planes of the face-centered cubic (FCC) lattice form the side facets of Au NWs. The Fe layers deposited on the {100} and {111} facets were found to be single-crystalline (SX) and poly-crystalline (PX), respectively.

Representative as-prepared bimetallic NWs with SX and PX Fe layers are presented in Fig. S1. The SX Fe layer on Au {100} facet exhibits a clear Bain orientation relationship with Au NW, e.g. [011]$_{Au}$//[001]$_{Fe}$, (011)$_{Au}$//(010)$_{Fe}$ and (100)$_{Au}$//(100)$_{Fe}$. The SX Fe-Au interface is without any misfit dislocations, as can be seen in the high-angle annular dark-field (HAADF) scanning transmission electron microscopy (STEM) image (Fig. S1e). This is because the lattice mismatch between (011)$_{Au}$//(010)$_{Fe}$ is only 0.59%, indicating the SX Fe-Au interfaces are coherent. The PX Fe layer is composed of Fe nano-grains with low misorientation angles, as indicated by its rough surface (Fig. S1f) and diffuse diffraction streaks in the diffraction patterns (Fig. 1d, h). Moreover, even though in some regions (110)$_{Fe}$ was found to be nearly parallel to (111)$_{Au}$, neither Nishiyama-Wasserman ([011]$_{Au}$//[001]$_{Fe}$) nor Kurdjumov-Sachs ([011]$_{Au}$//[111]$_{Fe}$) orientation relationships were observed (Fig. 1j). Therefore, we can conclude that the PX Fe-Au interfaces are of incoherent type.

The fact that coherent SX interface and incoherent PX interface were formed on {100} and {111} Au facets, respectively, was possibly due to the fact that the energy of the coherent interface is significantly lower than that of its incoherent counterpart [48]. This resulted in Frank-van-der-Merwe growth of Fe on {100} facets, and in Volmer-Weber growth on {111} facets of Au NW [36].

**Acknowledgements**
This work was supported by German-Israeli Foundation for Scientific Research and Development, Grant No. I-1360-401.10/2016, and by the Russell Berrie Nanotechnology Institute at the Technion. Helpful discussions with Dr. Leonid Klinger are heartily appreciated.

**Author contributions:** G.R. and E.R initiated and directed the study. G.R. and E.S. synthesized the NWs. Y.Q. carried out the experiments and analyzed the data. M.K. took part in *in-situ* heating experiments. Y.Q. and E.R. wrote the manuscript, with inputs from all co-authors.

**Competing interests:** All authors declare no competing interests.

Supplementary Materials for

# Plastic forming of metals at the nanoscale: interdiffusion-induced bending of bimetallic nanowhiskers

Yuanshen Qi[1*], Gunther Richter[2], Eylül Suadiye[2], Michael Kalina[1], Eugen Rabkin[1*]

[1]Department of Materials Science and Engineering, Technion – Israel Institute of Technology, 3200003 Haifa, Israel

[2]Max Planck Institute for Intelligent Systems, Heisenbergstrasse 3, 70569 Stuttgart, Germany

*Correspondence to: yuanshen.qi@campus.technion.ac.il; erabkin@technion.ac.il

Address: Department of Materials Science and Engineering, Technion – Israel Institute of Technology, 3200003 Haifa, Israel

## Additional results:

Estimating the bending of the NW#1 at 25°C due to lattice mismatch strain and island-zipping stress

We consider two physical reasons for the observed bending in the as-synthesized NW#1 at 25 °C, i.e. the lattice mismatch strain at the coherent single-crystalline (SX) Fe-Au interface, and the strain induced by the island-coalescence stress developed in the poly-crystalline (PX) Fe layer.

Let us first estimate the contribution from the lattice mismatch strain. Because there is no coherency strain in the PX Fe layer adjacent to the incoherent PX Fe-Au interface, we will only consider the lattice mismatch strain at the coherent SX Fe-Au interface. Here, we use a simplified model, see Figure. S2, in which the Au NW has a square cross-section with four {100} facets, with the SX Fe layer covering one facet in the Bain orientation relationship with the Au NW. The mismatch strain between $(011)_{Au}//(010)_{Fe}$ on the Au NW can be estimated as $\varepsilon_m = d_{Fe(010)} - d_{Au(011)} / d_{Au(011)} = -0.0059$, where $d_{Fe(010)}$ =0.2867 nm and $d_{Au(011)}$=0.2884 nm are the interplanar spacings in Fe and Au, respectively. To estimate the the radius of curvature, $R$, the Timoshenko formula [1] describing the bending of bimetallic beams is employed here:

$$R = \frac{E_{Fe}^2 h_{Fe}^4 + 4 E_{Fe} E_{Au} h_{Fe}^3 h_{Au} + 6 E_{Fe} E_{Au} h_{Fe}^2 h_{Au}^2 + 4 E_{Fe} E_{Au} h_{Au}^3 h_{Fe} + E_{Au}^2 h_{Au}^4}{6 E_{Fe} E_{Au} (h_{Fe} + h_{Au}) h_{Fe} h_{Au} \cdot \varepsilon_m} \qquad (S1)$$

where $E_{Fe}$ =125 GPa and $E_{Au}$ = 82 GPa are the Young's moduli of Fe and Au along [001] and [011] crystallographic directions, respectively. $h_{Fe}$ = 160 nm and $h_{Au}$ = 200 nm are the thicknesses of the Fe and Au constituents, respectively, estimated from Figure S2 (d).

The radius of curvature, $R$, is estimated to be 40 µm, which is somewhat larger than the measured one, 23±1 µm, in Figure 1. This difference could be caused by the tensile stress developed in the PX Fe layer during growth, and compressive strain in the Au NW in the vicinity of the interface, see Figure S3. It is known that during the Volmer–Weber film growth, the islands coalescence process generates a tensile stress which peaks during the islands coalescence and decreases gradually with increasing thickness of continuous film. Here we estimate the average tensile stress, $\langle \sigma \rangle$, using Nix-Clemens model [2]:

$$\langle \sigma \rangle = \left[ \left( \frac{1+v}{1-v} \right) \cdot E \cdot \frac{(2\gamma_s - \gamma_{gb})}{r} \right]^{1/2} \qquad (S2)$$

where $v$, $E$, $\gamma_s$, and $\gamma_{gb}$, are the Poisson's ratio, the Young's modulus, the surface energy and the grain boundary energy of ferrite α-Fe, which were chosen to be 0.3, 210 GPa, 2.1 J/m² and 1 J/m² [3], respectively. $r$ denotes the radius of the island and its chosen value is 33 nm, based on the measurements in Figure S2(e). The average tensile stress, $\langle \sigma \rangle$, is estimated to be 19.1 GPa. It is worth noting that this is a maximum stress reached at the moment of islands coalescence and formation of GBs. With increasing thickness, the island-coalescence tensile stress relaxes and the stress in the growing film becomes compressive due to adatoms penetration into the GBs [4]. Our results indicate that the internal stress in the PX Fe layer is tensile, and contributes to the bending of the as-prepared NW.

Estimating the bending of the NW#1 at 300 °C due to the mismatch of thermal expansion coefficients

The radius of curvature of NW#1 changed from 23±1 µm to 18±1 µm upon heating to 300 °C, which indicates an increase of the tensile stress in Fe coating. We attributed this phenomenon to the mismatch in the coefficients of thermal expansion of Au ($14.2 \cdot 10^{-6}$ °C$^{-1}$) and Fe ($11.4 \cdot 10^{-6}$ °C$^{-1}$). This mismatch resulted in a thermal strain of $-0.0008$ upon heating the NW#1 from room temperature (25 °C) to 300 °C. This thermal strain has to be added to the lattice mismatch strain of $-0.0059$, resulting in a total mismatch strain of $-0.0067$. According to Eq. (S1) this increase of strain results in the decrease of the NW radius of curvature from 40 µm to 36 µm, in qualitative agreement with the experimental data.

Kinetics of the nano-cavities healing in the PX Fe layer at 300 °C

We observed that while the radius of curvature of the NW#1 decreased to R = 18±1 µm upon heating to 300 ºC, it has not fully recovered to its original value $R = 23±1$ µm after the NW has been cooled down to 25 ºC. A slight decrease of the radius of curvature upon the full heating-cooling cycle (from 23±1 to 21±1 µm) is related with the irreversible processes occurring in the NW at elevated temperatures. We attribute this phenomenon to the healing of nano-cavities in the Fe layer, which results in a volume shrinkage of the Fe constituent and concomitant increase of the tensile stress in the layer. The healing and annihilation of the nano-cavities at elevated temperatures was demonstrated in the ex-situ heat treatment, see Figs S4 and S5. We assume that the healing process occurred via Fe atoms diffusion through a GB connected to the nano-cavity. The GB itself plays a role of a source of Fe atoms (or a sink of vacancies), which in turn contributes to the increase of tensile stress in the Fe coating. To prove this hypothesis, we propose a model to estimate the time required for Fe GB diffusion to heal a slit-shaped nano-cavity. In our model, the main driving force for the nano-cavity healing is the decrease of energy associated with replacing two free Fe surfaces (side surfaces of the slit) with an Fe GB.

Let us consider a slit-shaped nano-cavity with the thickness $h=3$ nm and total length $L_0= 50$ nm, located at the distance $l_0 = 20$ nm away from the Fe-Au interface (Fig. S6). A GB in the PX Fe layer is connected to the tip of the nano-cavity. The depletion of Fe atoms at the GB and their diffusion towards the nano-cavity lead to the latter retraction and increase of the distance between the nano-cavity tip and the Fe-Au interface, $l$. We will assume that the Fe grains on both sides of the GB homogeneously drift towards the GB (this way, the generation of additional elastic stresses at the GB associated with normal displacement variations is avoided). In this case, the distribution of chemical potential of Fe atoms along the GB, $\mu(x)$, is a parabolic (second-order) function of the distance, $x$, from the Fe-Au interface, $\mu(x)=ax^2+bx+c$ [5]. Furthermore, we assume that there is no Fe diffusion flux entering the GB at the Fe-Au interface, $\left.\frac{\partial \mu}{\partial x}\right|_{x=0} = 0$. This boundary condition results in $b = 0$ and $\mu(x)=ax^2+c$. The chemical potential of Fe atoms averaged over the whole GB length, $l$, is

$$\bar{\mu}_0 = \frac{1}{l}\int_0^l (ax^2 + c)dx = \frac{1}{l}\left(\frac{al^3}{3} + cl\right) \tag{S3}$$

On the other hand, the average chemical potential of Fe atoms that is $l$ away from the Fe-Au interface can be estimated by adding an infinitesimally thin layer of Fe at the GB, calculating the change of the total energy of the system, and normalizing it by the number of added atoms:

$$\bar{\mu}_0 = \Omega\frac{\gamma_s + \gamma_i - \gamma_{Au}}{l} \tag{S4}$$

where $\Omega$ denotes the atomic volume of Fe; $\gamma_s$, $\gamma_i$, and $\gamma_{Au}$ are the surface energy of Fe, Fe/Au interface energy and surface energy of Au, respectively.

Substituting Eq. (S3) in Eq. (S4) yields

$$\frac{al^2}{3} + c = \Omega\frac{\gamma_s + \gamma_i - \gamma_{Au}}{l} \tag{S5}$$

Furthermore, the chemical potential of the Fe atoms on the inner surface of the nano-cavity tip (i.e. at the distance $l$ from the Fe-Au interface) is

$$\mu_f = al^2 + c \approx \Omega\frac{\gamma_{gb} - 2\gamma_s}{h} \tag{S6}$$

where $\gamma_{gb}$ is the GB energy of ferrite α-Fe. The last term on the right hand side of Eq. (S6) is associated with the curvature of the inner surface of the nano-cavity at the tip. Solving Eqs (S5-S6) yields

$$a = \frac{3\Omega}{2l^2}\left(\frac{\gamma_{gb} - 2\gamma_s}{h} - \frac{\gamma_s + \gamma_i - \gamma_{Au}}{l}\right) \tag{S7}$$

$$c = \frac{\Omega}{2}\left(3\cdot\frac{\gamma_s + \gamma_i - \gamma_{Au}}{l} - \frac{\gamma_{gb} - 2\gamma_s}{h}\right) \tag{S8}$$

Then the GB diffusion flux of Fe atoms entering the nano-cavity at the intersection line of the GB and the inner surface of the nano-cavity is

$$J = -\frac{D_{gb}\delta}{kT}\cdot\frac{\partial\mu}{\partial x}\bigg|_{x=l} = -\frac{D_{gb}\delta}{kT}\cdot\frac{3\Omega}{l}\left(\frac{\gamma_{gb} - 2\gamma_s}{h} - \frac{\gamma_s + \gamma_i - \gamma_{Au}}{l}\right) \tag{S9}$$

where $D_{gb}$ and $\delta$ are the self-diffusion coefficient of Fe along the GB at 300 °C and the GB width, respectively. $kT$ has its usual thermodynamic meaning. Assuming the constant width of the nano-cavity, this flux cases the nano-cavity retraction according to

$$J = h\cdot\frac{dl}{dt} \tag{S10}$$

Substituting Eq. (S10) in Eq. (S9) yields

$$\frac{dl}{dt} = \frac{3D_{gb}\delta\Omega}{kTlh}\left(\frac{2\gamma_s - \gamma_{gb}}{h} + \frac{\gamma_s + \gamma_i - \gamma_{Au}}{l}\right) \tag{S11}$$

For $l_0 \gg h$ the magnitude of the second term on the RHS of Eq. (S11) is small compared to the first term, and the Eq. (S11) can be approximately re-written as

$$\frac{dl^2}{dt} \approx \frac{6D_{gb}\delta\Omega}{kTh^2}(2\gamma_s - \gamma_{gb}) \tag{S12}$$

Solving this differential equation yields the following solution for the annealing time required for nano-cavity retraction by the distance $l-l_0$:

$$l^2 - l_0^2 \approx \frac{6D_{gb}\delta\Omega \cdot (2\gamma_s - \gamma_{gb})}{kTh^2}t \tag{S13}$$

where $l$, $l_0$, $T$, $h$, $\delta$, $\Omega$, $\gamma_s$ and $\gamma_{gb}$ are chosen to be 70 nm, 20 nm, 573 K, 3 nm, 0.5 nm, 7.1·10$^{-6}$ m$^3$/mol, 2.1 J/m$^2$, and 1 J/m$^2$, respectively. $D_{gb}$ was chosen in the range between 1.3·10$^{-18}$ m$^2$/s [6] and 2.8·10$^{-16}$ m$^2$/s [7]. The time of nano-cavity retraction estimated with the aid of Eq. (S13) is then in the range from 17 s to 1 h, depending on the GB diffusivity, in good agreement with our experimental observations. Therefore, the nano-cavities healing at the temperature of 300 °C leading to the increase of tensile stress in the Fe coating is feasible on the time scale of our experiment.

Formation of a tilt GB at the mouth of Au layer penetrating into the gap between PX and SX Fe layers

We will consider a simple model of a rectangular gap between two Fe layers which is filled with Au during high-temperature annealing (Fig. S8). We will neglect the changes in the geometry of Au NW upon the gap filling, since the total volume of Au penetrating the gap is negligible in comparison to the volume of the Au NW. Thus, the total energy of all surfaces and interfaces only in the region of the gap will be considered. The initial energy per unit of longitudinal length, $E_{total}^{25°C}$, can be written as

$$E_{total}^{25°C} = L(\gamma_{S-Fe}^1 + \gamma_{S-Fe}^2) + h\gamma_{S-Au}^1 \tag{S14}$$

where $L$ and $h$ are the length and the width of the gap, respectively. $\gamma_{S-Fe}^1$ and $\gamma_{S-Fe}^2$ are the average surface energy of the PX Fe layer, and the energy of the side surface of the SX Fe layer, respectively. $\gamma_{S-Au}^1$ is the energy of exposed Au surface at the entrance to the gap. Assuming that the gap is filled with Au in homoepitaxial orientation relationship with the underlying Au NW yields the following expression for the energy of the filled gap:

$$E_{total}^{500°C} = L(\gamma_{In-Fe}^1 + \gamma_{In-Fe}^2) + h\gamma_{S-Au}^1 \tag{S15}$$

where $\gamma_{In-Fe}^1$ and $\gamma_{In-Fe}^2$ are the average energy of the PX Fe-Au interface, and the energy of the SX Fe-(277) Au interface, respectively. Another possibility is for the Au layer penetrating the gap to change its orientation, so that a low-energy SX Fe-(100) Au interface is formed on the right side of the gap. In this case, the total energy of the filled gap will be different from the one given by Eq. (S15):

$$E_{total}^{500°C*} = L\left(\gamma_{In-Fe}^{1*} + \gamma_{In-Fe}^{2*}\right) + h\left(\gamma_{S-Au}^{1*} + \gamma_{GB-Au}^{*}\right) \tag{S16}$$

where $\gamma_{GB-Au}^{*}$ is the energy of the GB formed at the mouth of the Au layer, and the star superscript denotes the change of surface orientation or interface crystallography. For simplicity, we assumed that the cross-sectional length of the newly formed GB is the same as the gap width $h$. Since the anisotropy of surface energy of cubic metals rarely exceeds 2-3%, we will assume $\gamma_{S-Fe}^{1} \approx \gamma_{S-Fe}^{2} \approx \gamma_s \approx 2.1$ J/m$^2$, and $\gamma_{S-Au}^{1} \approx \gamma_{S-Au}^{1*} \approx \gamma_{Au} \approx 1.4$ J/m$^2$. Because the average energy of the PX Fe-Au interface is determined for many different orientations of the Fe grains, we will assume the equal values of $\gamma_{In-Fe}^{1}$, $\gamma_{In-Fe}^{1*}$, and of the incoherent interface energy, $\gamma_{In-Fe}^{2}$. We will estimate the values of these energies by the energy value of high-angle GB in Fe, ~1 J/m$^2$ [3]. The density functional theory [8] and regular solution model [9] – based estimates of the energy of the coherent Fe-Au interface, $\gamma_{In-Fe}^{2*}$, yield the values in the range of 0.36-0.5 J/m$^2$. Finally, the energy of high-angle GB in Au will be estimated as one-third of its surface energy, $\gamma_{GB-Au}^{*} \approx 0.5$ J/m$^2$. Comparing the Eqs (S14)-(S16) reveals that filling the gap between the two Fe layers by Au is energetically favorable for any values of $L$ and $h$. Moreover, for $h < (1-1.3) \cdot L$, the formation of GB by lattice rotation about $[0\bar{1}1]_{Au}$ axis leading to establishing a coherency at the SX Fe-Au interface becomes energetically more favorable than homoepitaxial penetration of Au into the gap. The analysis of Fig. 4c, d in the main text and Fig. S8 demonstrates that $h$ (<10 nm) << $L$ (50-200 nm) and, therefore, the formation of a GB is energetically favorable.

Estimation of the GB Kirkendall effect induced irreversible bending at the temperatures above 300 °C

The NW#1 bent in the opposite direction at 400 °C and 500 °C, indicating that the stress state of the Fe layer changed from tension to compression. We attribute this lateral expansion of the PX Fe layer to the accretion of Au atoms at the GBs in Fe, as observed in the STEM-EDX micrographs, and is shown schematically in Fig. S10. However, due to the high solubility of Fe in Au [10] the diffusion flux of Au along the GBs in Fe can be accompanied by nearly equal flux of Fe atoms in the opposite direction, leaving the Fe layer and dissolving in the Au NW (chemical GB interdiffusion). Thus, Fe atoms leaving the Fe layer may compensate its lateral expansion due to accretion of Au atoms. We will estimate below whether a simple GB interdiffusion with equal partial GB diffusion coefficients of Au and Fe (i.e. full balance of the GB diffusion fluxes) can account for the observed bending of the NW#1.

The diameters of the Fe and Au atoms are 0.2482 and 0.2884 nm, respectively. With the average grain size in the PX Fe layer of 66 nm, replacing one monolayer of Fe at the GB with a monolayer of Au will cause a lateral strain of 6×10$^{-4}$, one order of magnitude lower than the initial mismatch strain. Therefore, the simple GB interdiffusion with equal partial GB diffusion coefficients of Au and Fe cannot account for the NW bending in the opposite direction. Let us now assume that Au diffuses much faster than Fe along the Fe GBs (GB Kirkendall effect). In this case, accretion of Au at the GBs will not be accompanied by any significant outdiffusion of Fe. One full monolayer of Au will cause a strain of 0.004, comparable with the lattice mismatch strain. Accommodating two-three monolayers of Au in the Fe GBs will fully compensate the lattice mismatch strain and lead to the NW bending in the opposite direction, in accordance with our experimental observations. From the measurements in Fig. S7b, the width of the Au-rich diffusion/segregation layer at the Fe GBs after annealing at 400 °C and 500 °C for 20 min is about 2 nm. Thus we conclude that the major part of plastic bending of the bi-metallic Au-Fe NW at high temperatures is caused by the GB Kirkendall effect. It should be noted that the formation of "GB diffusion wedges" of the diffuser during the GB Kirkendall effect has been described by Klinger and Rabkin [11].

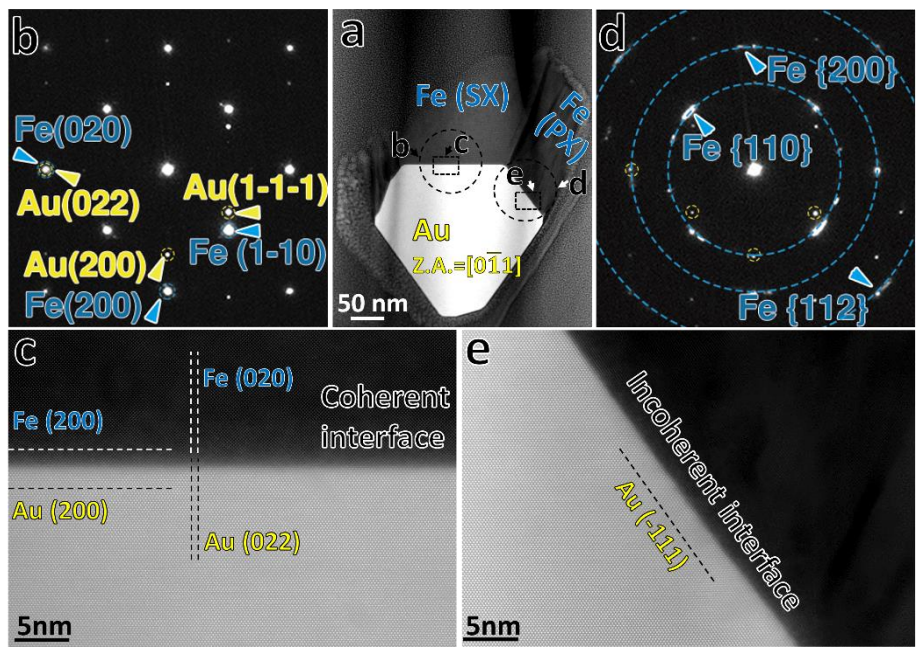

Figure S1. Interfaces in bimetallic Au/Fe NW. (a) HAADF-STEM image of the cross-section of a representative NW; (b, c) the SADP and atomic-resolution HAADF-STEM image showing a single-crystalline Fe layer on a Au (100) facet and a coherent interface between them; (d, e) the SADP and atomic-resolution HAADF-STEM images showing a poly-crystalline Fe layer on a Au (111) facet and an incoherent interface between them. Images were taken along the [0$\bar{1}$1] Z.A. of Au.

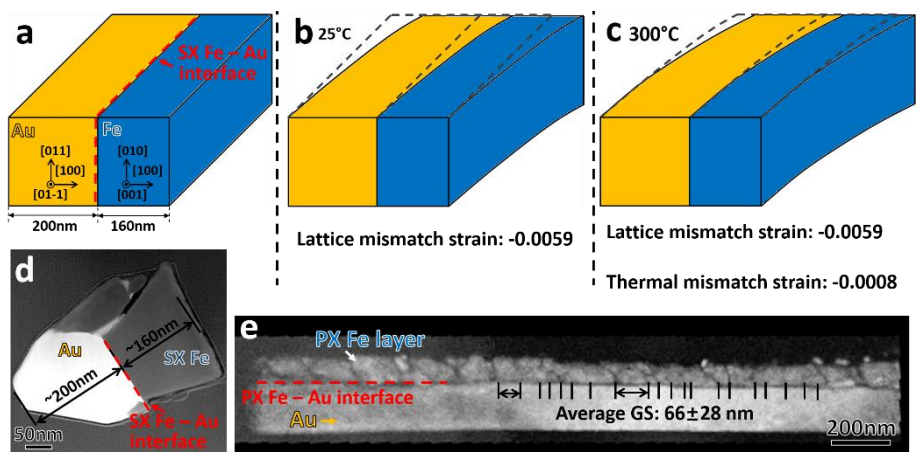

Figure S2. Illustration of the initial NW#1 bending due to lattice mismatch strain and island coalescence stress. (a-d) the lattice mismatch strain induced by the heteroepitaxy at the single-crystalline Fe – Au interface; (a) the model of initial configuration; (b) the bending induced by the lattice mismatch strain; (c) additional elastic bending at 300 °C due to the mismatch in thermal expansion coefficients of Au and Fe; (d) the geometric parameters used for the estimation of the NW radius of curvature; (e) Longitudinal cross-sectional TKD band contrast image of the NW#1 illustrating the geometric parameters used for the estimation of island coalescence stress in PX Fe layer.

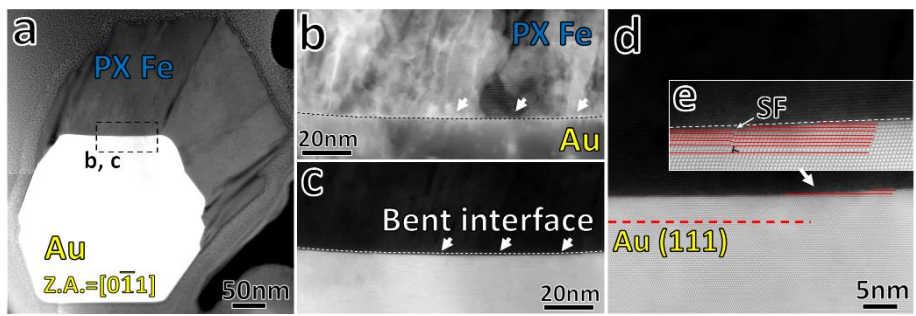

Figure S3. A demonstration of Fe-Au interface bending due to island coalescence in the PX Fe layer. (a) HAADF-STEM of the cross-section of the Au/Fe NW; (b, c) LAADF and HAADF-STEM images showing the bending of the PX Fe layer – Au interface due to the island coalescence induced tensile stress in the Fe layer; arrows and dashed lines highlight the bent interface; (d, e) atomic-resolution HAADF-STEM image showing the ledges on the interface and a stacking fault (SF) in the Au NW. The formation of interface ledges and a Shockley partial dislocation indicated the strain was partially relieved by plastic deformation. Images were taken along the [0$\bar{1}$1] Z.A. of Au.

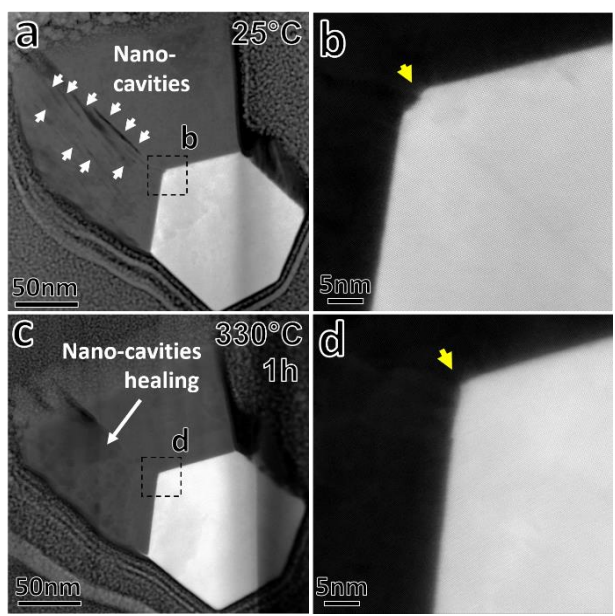

Figure S4 Quasi-in-situ demonstration of nano-cavities healing after annealing. HAADF-STEM images (a, b) showing the cross-section of a bimetallic NW in which some nano-cavities in the Fe layer were highlighted by arrows; (c, d) the same NW after annealing at 330 °C for 60 min in RTA exhibited much fewer nano-cavities. The atomic resolution HAADF-STEM images of (b, d) demonstrate that the Au-Fe interface at the facets intersection migrated towards Fe after annealing. Images were taken along the [0$\bar{1}$1] Z.A. of Au.

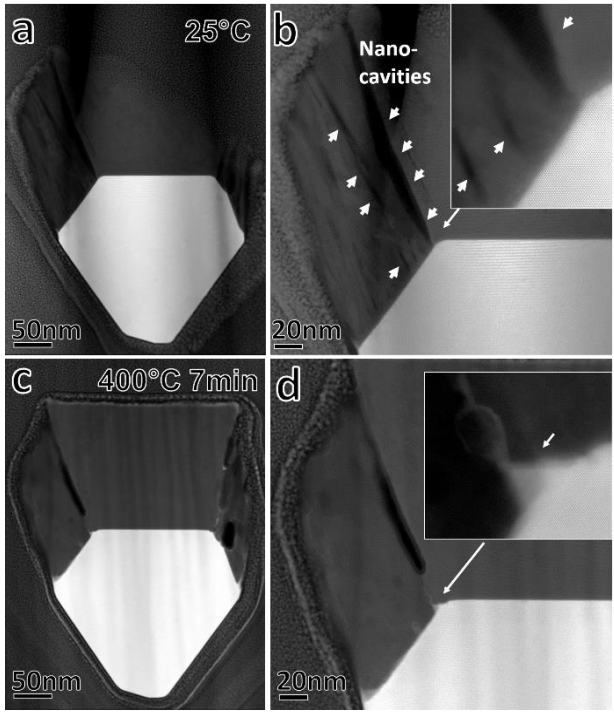

Figure S5. Quasi-in-situ demonstration of nano-cavities healing and Au penetration into Fe after annealing. HAADF-STEM images (a, b) showing the cross-section of a bimetallic NW in which some nano-cavities in the Fe layer were highlighted by arrows; (c, d) the same NW after annealing at 400 °C for 7 min in RTA exhibited much fewer nano-cavities. The enlarged HAADF-STEM images of (b, d) demonstrate the diffusion and penetration of Au into the nano-cavity at the intersection of Au facets. Images were taken along the $[0\bar{1}1]$ Z.A. of Au.

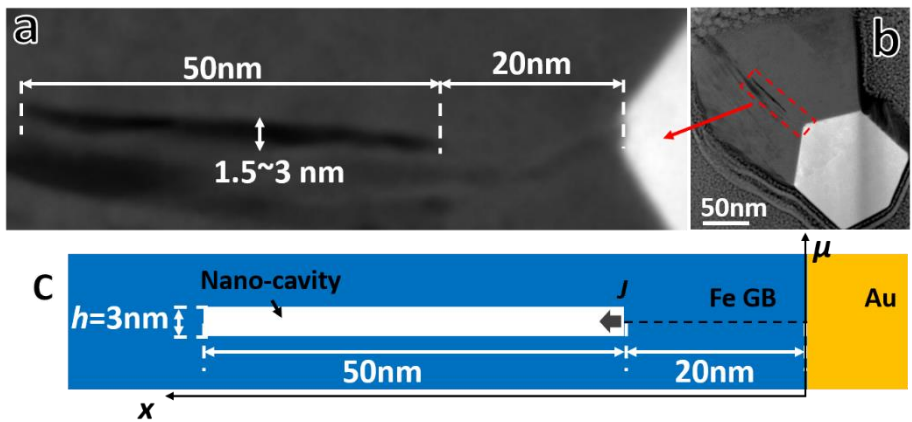

Figure S6. The geometric parameters of the model of a Fe nano-cavity healing by grain boundary (GB) diffusion mechanism. (a, b) HAADF-STEM images showing a representative nano-cavity in the Fe layers; (c) the schematic illustration of the nano-cavity.

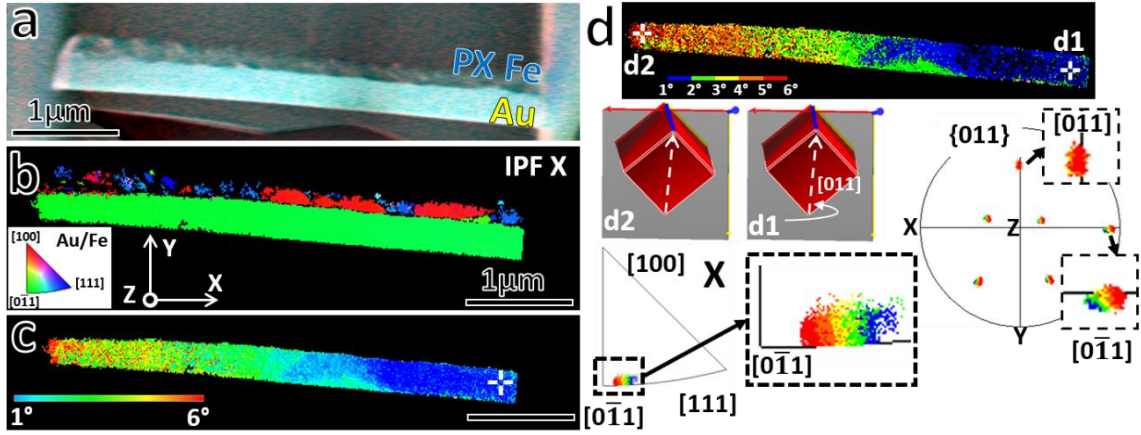

Figure S7. Lattice distortion in bent NW#1 characterized by TKD. (a) the color-coded dark-field (CCDF) image of the longitudinal cross-section of NW#1; (b) the orientation image with an insert illustrating the color codes of normal (Z) grain orientations; (c) the misorientation map of the Au NW where the reference location is pointed by the white cross. The legend shows the range of misorientations (1°-6°); (d) segmented misorientation map, inverse pole figure and pole figure showing the Au lattice distortion was caused by the lattice rotation around $[10\bar{1}]$ zone for 6°.

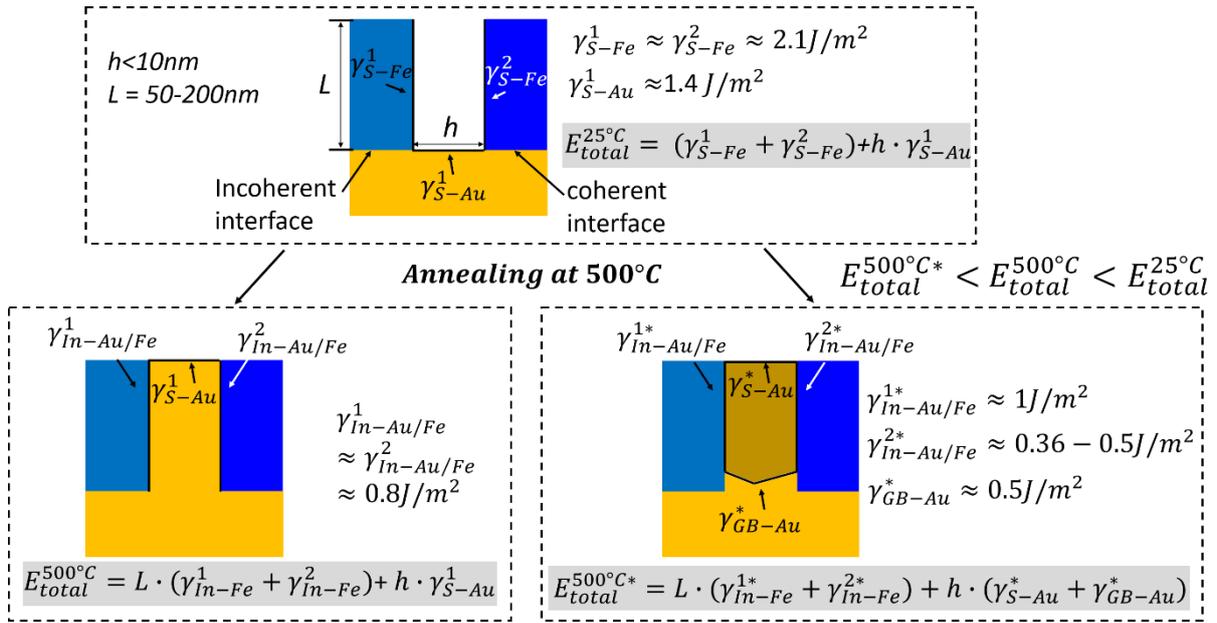

Figure S8. Energy minimization approach employed to explain the formation of tilt GB at the "mouth" of Au layer penetrating into the Fe nano-cavity.

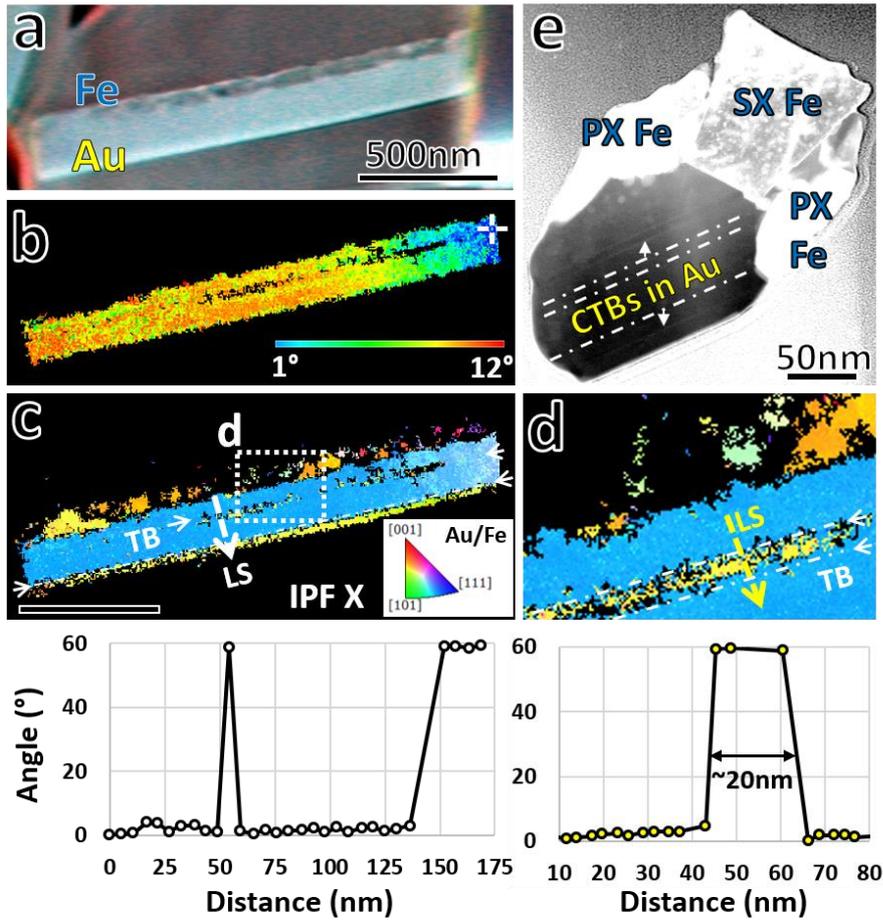

Figure S9 Lattice distortion and deformation twinning in bent NW#2 characterized by TKD and LAADF. (a) the CCDF image of the longitudinal cross-section of NW#2; (b) the misorientation map of the Au NW where the reference location is pointed by the white cross. The legend shows the range of misorientations (1°-12°); (c, d) the orientation images showing the twin boundary (TB) and misorientation line profile from the line scan (LS); (e) LAADF image showing the twin boundaries in the cross-section normal to the NW axis $[0\bar{1}1]_{Au}$.

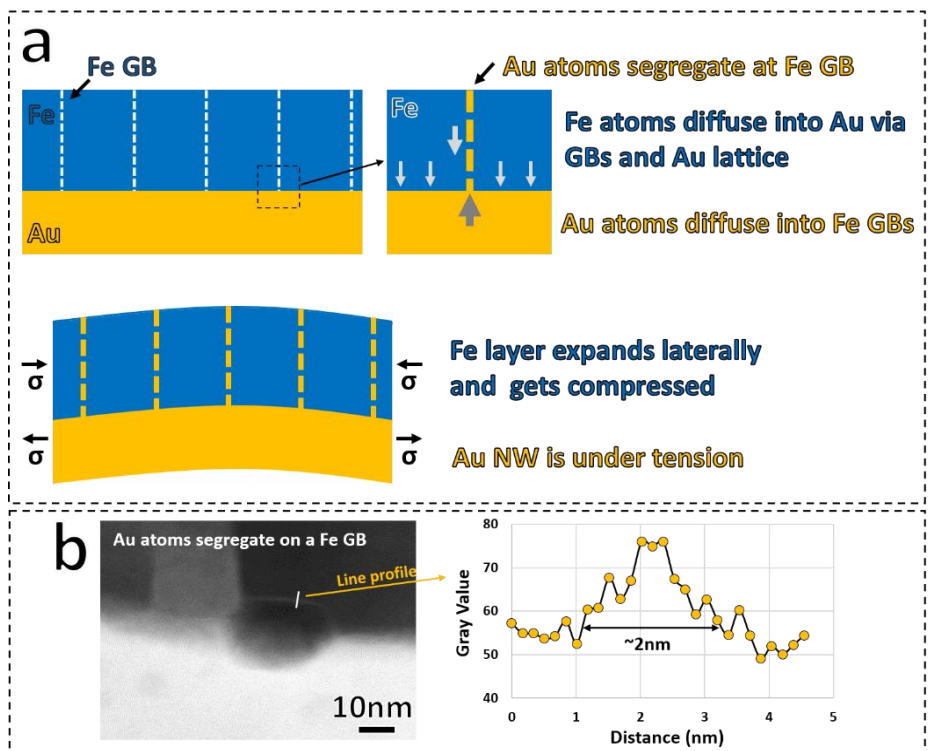

Figure S10. Au atoms accretion at the GBs in Fe causing lateral expansion of the Fe layer and the NW bending towards Au. (a) schematic illustration showing the interdiffusion between Au NW and PX Fe layer at elevated temperature; (b) the HAADF-STEM image-based measurement of the width of Au diffusion/segregation layer at the Fe GB after the in-situ annealing.

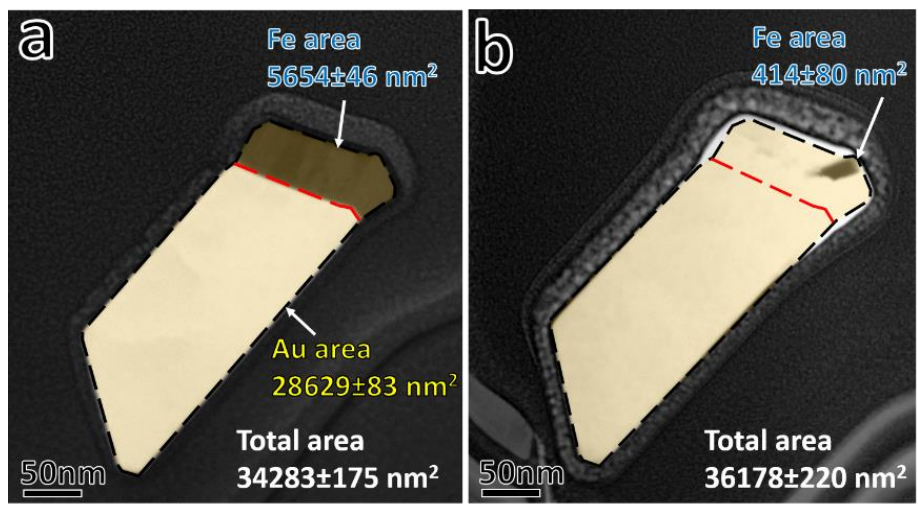

Figure S11. The evolution of the cross-sectional dimensions of the NW#3 after annealing at 600 °C for 30 min. The estimation was based on the measurements of the areas in HAADF-STEM images before (a) and after (b) annealing. Images were taken along the [0$\bar{1}$1] Z.A. of Au.

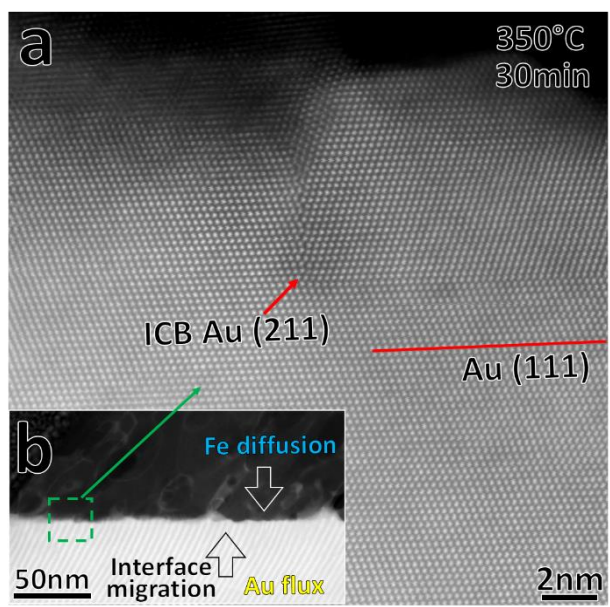

Figure S12. The formation of an incoherent twin boundary (ITB) induced by interface migration and bulk Fe-Au interdiffusion. (a) an atomic resolution HAADF-STEM image showing an ITB (211) formed in the vicinity of the incoherent Au – PX Fe interface after annealing at 350 °C for 30 min. (b) the lower magnification HAADF-STEM image showing the nano-roughness of the interface which indicates the migration process. Images were taken along the $[0\bar{1}1]$ Z.A. of Au.